\documentclass[12pt]{article}

\usepackage{graphicx}
\begin{document}

%New Commands
\newcommand{\siml}{\stackrel{<}{\sim}}
\newcommand{\simg}{\stackrel{>}{\sim}}
\newcommand{\lleq}{\stackrel{<}{=}}

\baselineskip=1.333\baselineskip
%PRE
%\baselineskip=2.0\baselineskip

%\draft
%\baselineskip=0.5\baselineskip

%
\begin{center}
{\large\bf
Dynamics of the Langevin model subjected to colored noise:
Functional-integral method 
%\footnote{e-print arXiv: 0708.2563}
} 
\end{center}

\begin{center}
Hideo Hasegawa
\footnote{e-mail address:  hideohasegawa@goo.jp}
\end{center}

\begin{center}
{\it Department of Physics, Tokyo Gakugei University  \\
Koganei, Tokyo 184-8501, Japan}
\end{center}
\begin{center}
%{\rm (Jan. 11, 2005)}
({\today})
\end{center}
%\maketitle
\thispagestyle{myheadings}

\begin{abstract}
We have discussed the dynamics of Langevin model 
subjected to colored noise, by using the functional-integral 
method (FIM) combined with equations of motion for mean and 
variance of the state variable.
Two sets of colored noise have been investigated:
(a) one additive and one multiplicative colored noise, 
and (b) one additive and two multiplicative colored noise. 
The case (b) is examined with the relevance to a recent controversy
on the stationary subthreshold voltage distribution
of an integrate-and-fire model including stochastic
excitatory and inhibitory synapses and a noisy input.
We have studied the stationary probability distribution 
and dynamical responses to time-dependent (pulse and sinusoidal) 
inputs of the linear Langevin model.
Model calculations have shown that results of the FIM are 
in good agreement with those of direct simulations (DSs).
A comparison is made among various approximate analytic
solutions such as the universal colored noise approximation (UCNA).
It has been pointed out that dynamical responses to pulse and sinusoidal 
inputs calculated by the UCNA are rather different from those 
of DS and the FIM, although they yield the same 
stationary distribution.  

\end{abstract}

%\noindent
\vspace{0.5cm}


\vspace{0.5cm}


\vspace{0.5cm}

{\it Address:} 4-1-1, Nukui-kita machi, Koganei, Tokyo 184-8501, Japan

\newpage
%\narrowtext
\section{Introduction}

Nonlinear stochastic dynamics of physical, chemical, biological
and economical systems has been extensively studied
(for a recent review, see Ref. \cite{Lindner04}).
In most theoretical studies, Gaussian white noise
is employed as random driving force because of its
mathematical simplicity.
The white-noise approximation is appropriate to systems
in which the time scale characterizing the relaxation of the noise
is much shorter than the characteristic time scale of the system.
There has been a growing interest in theoretical study of nonlinear
dynamical systems subjected to colored noise 
with the finite correlation time
(for a review on colored noise, see Ref. \cite{Hanggi95}:
related references therein).
It has been realized 
that colored noise gives rise to
new intriguing effects such as the reentrant phenomenon 
in a noise-induced transition \cite{Reent}
and a resonant activation in bistable systems
\cite{RA}.

The original model for a system driven by colored noise
is expressed by non-Markovian stochastic differential equation.
This problem may be transformed to a Markovian one, by extending
the number of relevant variables and including an additional 
differential equation describing the Orstein-Uhlenbeck (OU) process.
It is difficult to analytically solve the Langevin model
subjected to {\it colored} noise.
For its analytical study, two approaches have been adopted:
(1) to construct the multi-dimensional Fokker-Planck equation (FPE) 
for the multivariate probability distribution, and
(2) to derive the effective one-dimensional FPE equation.
The presence of multi-variables
in the approach (1) makes a calculation of even the stationary
distribution much difficult.
In a recent study on the Langevin model subjected to additive (non-Gaussian) 
colored noise \cite{Hasegawa07b}, we employed the approach (1),
analyzing the multivariate FPE 
with the use of the second-order moment method.  
A typical example of the approach (2) is the universal colored
noise approximation (UCNA) \cite{Jung87},
which interpolates between the limits of zero and infinite
relaxation times, and
which has been widely adopted for a study of colored noise \cite{Hanggi95}.
Another example of the approach (2)
is the path-integral and functional-integral methods 
\cite{Sancho82}-\cite{Wu07}
obtaining the effective FPE, with which stationary properties
such as the non-Gaussian stationary distribution
have been studied \cite{Hanggi95}.

Theoretical study on the Langevin model driven by colored noise
has been mostly made for its stationary properties 
such as the stationary probability distribution 
and the phase diagram of noise-induced transition
\cite{Hanggi95}.
As far as we are aware of, little theoretical study has
been reported on dynamical properties 
such as the response to time-dependent inputs.
Refs. \cite{Brunel01,Fourcaud02} have discussed
the filtering effect, in which the high-frequency response 
of the system is shown to be improved by colored noise.
%except for stochastic resonance and resonant activation.
The purpose of the present paper is to extend the 
functional-integral method (FIM)
such that we may discuss dynamical properties of
the Langevin model subjected to colored noise.
We consider, in this paper, two sets of colored noise:
(a) one additive and one
multiplicative colored noise, and (b) one additive and 
two multiplicative colored noise. 
The case (b) is included to clarify, to some extent, 
a recent controversy
on the subthreshold voltage distribution of a leaky 
integrate-and-fire model including 
conductance-based stochastic
excitatory and inhibitory synapses as well as noisy inputs
\cite{Rudolph03}-\cite{Rudolph06}. 

The paper is organized as follows.
The FIM is applied to the above-mentioned cases (a)
and (b) in Secs. 2 and 3, respectively, where
the stationary distribution and the response to
time-dependent inputs are studied.
In Sec. 4, we will discuss the recent controversy on
the subthreshold voltage distribution of a leaky 
integrate-and-fire model 
\cite{Rudolph03}-\cite{Rudolph06}. 
A comparison is made among results of
some approximate analytical theories
such as the UCNA  \cite{Hanggi95,Jung87}. 
The final Sec. 5 is devoted to conclusion.

%\newpage
\section{Langevin model subjected to one additive and
one multiplicative colored noise}
%\section{Case of one additive and
%one multiplicative colored noise}

\subsection{Effective Langevin equation}

We have considered the Langevin model subjected to
additive and multiplicative colored noise given by
\begin{eqnarray}
\frac{dx}{dt}\!\!&=&\!\!F(x) + \eta_0(t)
+ G(x) \eta_1(t), 
\end{eqnarray}
with
\begin{eqnarray}
\frac{d \eta_m(t)}{d t}
&=& -\frac{\eta_m}{\tau_m}+ \frac{\sqrt{2 D_m}}{\tau_m} \;\xi_m(t),
\hspace{1.0cm}\mbox{($m= 0$ and 1)} %\nonumber
\end{eqnarray}
where $F(x)$ and $G(x)$ denote arbitrary functions of $x$:
$\eta_0(t)$ and $\eta_1(t)$ stand for additive 
and multiplicative noise, respectively:
%$I(t)$ expresses an input signal:
$\tau_m$ and $D_m$ express the relaxation times
and the strengths of colored noise for additive ($m=0$) 
and multiplicative noise ($m=1$):
$\eta_m(t)$ express independent zero-mean Gaussian white
noise with correlations given by
\begin{eqnarray}
\langle \xi_m(t)\:\xi_n(t') \rangle
&=& \delta_{mn} \delta(t-t').
\end{eqnarray}
The distribution and correlation of $\eta_m$ are given by
\begin{eqnarray}
p(\eta_m) & \propto& \exp\left(-\frac{\tau_m}{2 D_m} \eta_m^2 \right),
\end{eqnarray}
\begin{eqnarray}
c_{mn}(t,t') &=&
\langle \eta_m(t) \eta_n(t')  \rangle 
=  \delta_{mn} \left( \frac{D_m}{\tau_m} \right) 
\exp\left( -\frac{\mid t-t' \mid}{\tau_m} \right).
\end{eqnarray}

By applying the FIM to the Langevin model
given by Eqs. (1) and (2), we obtain the effective FPE given by 
\cite{Sancho82,Liang04} (details being given in the Appendix)
\begin{eqnarray}
\frac{\partial}{\partial t}p(x,t) 
&=& - \frac{\partial }{\partial x} \tilde{F}(x)p(x,t)
+ \tilde{D}_0 \frac{\partial^2 }{\partial x^2} p(x,t)
+ \tilde{D}_1 \frac{\partial }{\partial x}G(x)
\frac{\partial }{\partial x}G(x) p(x,t),
\end{eqnarray}
from which the effective Langevin model is derived as
\begin{eqnarray}
\frac{dx}{dt}\!\!&=&\!\!\tilde{F}(x) 
+ \sqrt{2 \tilde{D}_0} \xi_0(t)
+ \sqrt{2 \tilde{D}_1} G(x) \xi_1(t), 
\end{eqnarray}
with
\begin{eqnarray}
\tilde{F} &=& F, \\
\tilde{D}_0 &=& \frac{D_0}{(1-\tau_0 \langle F' \rangle )}, \\
\tilde{D}_1 &=& \frac{D_1}
{[1-\tau_1 (\langle F' \rangle -\langle  F G'/G \rangle)]}.
%\hspace{1.0cm}\mbox{($i= 1, 2$)} %\nonumber
\end{eqnarray}
Here $F'=dF/dx$ and $G'=dG/dx$,
and the bracket $\langle \cdot \rangle $ expresses
the average over $p(x,t)$ to be discussed shorty [Eq. (11)].
It is noted that
we will temporally evaluated $\langle F' \rangle$ {\it etc.}
in order to discuss dynamics of the system,
while they are conventionally evaluated for the stationary 
value as $F'(x_s)$ {\it etc.} with 
$x_s \:=\langle x(t=\infty) \rangle$ \cite{Sancho82,Liang04}.

\subsection{Equations of motion for mean and variance}

With the use of the effective FPE given by Eq. (7),
an equation of motion for the average of $Q(x)$:
\begin{eqnarray}
\langle Q \rangle &=& \int Q(x) \: p(x,t)\: dx,
\end{eqnarray}
is given by \cite{Hasegawa07a}
\begin{eqnarray}
\frac{d \langle Q \rangle}{d t}
&=& \langle Q' \tilde{F} \rangle + \tilde{D}_0 \langle Q'' \rangle
+ \tilde{D}_1 \langle (Q' G)' G \rangle,
\end{eqnarray}
which yields (for $Q=x, x^2$)
\begin{eqnarray}
\frac{d \langle x \rangle}{d t}
&=& \langle \tilde{F} \rangle 
+ \tilde{D}_1 \langle G' G \rangle, \\
\frac{d \langle x^2 \rangle}{d t}
&=& 2 \langle x \tilde{F} \rangle + 2 \tilde{D}_0
+ 2 \tilde{D}_1 \langle G^2+ x G' G \rangle.
\end{eqnarray}
Mean ($\mu$) and variance ($\gamma$) are defined by
\begin{eqnarray}
\mu &=& \langle x \rangle, \\
\gamma &=& \langle x^2 \rangle- \langle x \rangle^2.
\end{eqnarray}
Expanding Eqs. (13) and (14) around the mean value of $\mu$,
and retaining up to the second order of $\langle (\delta x_i)^2 \rangle$,
we get equations of motion for $\mu$ and $\gamma$
expressed by \cite{Hasegawa07a}
\begin{eqnarray}
\frac{d \mu}{d t} &=& \tilde{f}_0+ \tilde{f}_2 \gamma 
+\tilde{D}_1 [g_0 g_1 + 3(g_1 g_2+g_0 g_3)\gamma], \\
\frac{d \gamma}{dt} &=& 2 \tilde{f}_1 \gamma 
+ 4 \tilde{D}_1\gamma (g_1^2+2 g_0 g_2) 
+ 2 \tilde{D}_1 g_0^2 + 2 \tilde{D}_0,
\end{eqnarray}
%\begin{eqnarray}
%\tilde{D}_0 &=& \frac{D_0}{1-\tau_0 f_1 }, \\
%\tilde{D}_1 &=& \frac{D_1}{1-\tau_1 [f_1-f_0 \;g_1/g_0]},
%\hspace{1.0cm}\mbox{($i= 1, 2$)} %\nonumber
%\end{eqnarray}
where $\tilde{f}_{\ell}=(1/\ell !) 
\partial^{\ell} \tilde{F}(\mu)/\partial x^{\ell}$
and $g_{\ell}=(1/\ell !) \partial^{\ell}G(\mu)/\partial x^{\ell}$.
It is noted that $\tilde{D}_0$ and $\tilde{D}_1$ 
in Eqs. (17) and (18) are given by Eqs. (9) and (10), respectively.

In the case of $F(x)=-\lambda x+I$ and $G(x)=x$ where
$\lambda$ and $I$ denote the relaxation rate and
an input, respectively, the FIM yields
equations of motion for $\mu$ and $\gamma$ given by
\begin{eqnarray}
\frac{d \mu}{d t} &=& -\lambda \mu 
+\tilde{D}_1 \mu+I, \\
\frac{d \gamma}{dt} &=& -2 \lambda \gamma 
+ 4 \tilde{D}_1\gamma  
+ 2 \tilde{D}_1 \mu^2 + 2 \tilde{D}_0,
\end{eqnarray}
with
\begin{eqnarray}
\tilde{D}_0 &=& \frac{D_0}{(1+\lambda \tau_0 ) }, \\
\tilde{D}_1 &=& \frac{D_1}{[1+(\tau_1I/\mu)]}.
%\hspace{1.0cm}\mbox{($i= 1, 2$)} %\nonumber
\end{eqnarray}
We have to solve Eqs. (19)-(22) 
for $\mu$, $\gamma$, $\tilde{D}_0$ and $\tilde{D}_1$
in a self-consistent way.
%: $\tilde{D}_1$ generally becomes time dependent.

Stationary values of $\mu$ and $\gamma$ are implicitly given by
\begin{eqnarray}
\mu_s &=& \frac{I}{(\lambda - \tilde{D}_1)}, \\
\gamma_s &=& \frac{(\tilde{D}_0+ \tilde{D}_1 \mu_s^2)}
{( \lambda - 2\tilde{D}_1) },
\end{eqnarray}
with $\tilde{D}_0$ and $\tilde{D}_1$ given by Eqs. (21) and (22),
respectively, with $\mu=\mu_s$.
Equations (23) and (24) show that $\mu_s$ and $\gamma_s$ diverge
for $\tilde{D}_1 > \lambda$ and $\tilde{D}_1 > \lambda/2$, respectively.
The divergence of moments is common in systems subjected
to multiplicative noise, because its stationary distribution 
has a long-tail power-law structure
\cite{Hasegawa07a}-\cite{Anten03}.
From Eqs. (22) and (23), we get
\begin{eqnarray}
\tilde{D}_1 &=& \frac{D_1}{[1+\tau_1(\lambda-\tilde{D}_1)]}, \\
&=& \frac{1}{2\tau_1}
\left[(1+\lambda \tau_1)-\sqrt{(1+\lambda \tau_1)^2-4 \tau_1 D_1}\right], 
\end{eqnarray}
%In the limit of $\tau_1 D_1/(1+\lambda \tau_1)^2 \ll 1$, 
which yields
\begin{eqnarray}
\tilde{D}_1 
& = & \left( \frac{D_1}{1+\lambda \tau_1} \right)
\left[ 1+\frac{\tau_1 D_1}{(1+\lambda \tau_1)^2}
-2 \frac{\tau_1^2D_1^2}{(1+\lambda \tau_1)^4} + \cdot\cdot \right],
\nonumber \\
& \simeq & \frac{D_1}{(1+\lambda \tau_1 )}
\equiv \tilde{D}_1^{APP}.
%\left[1 - \frac{\tau_1 D_1}{(1+\lambda \tau_1)^2} + \cdot \cdot \right], 
\hspace{1.0cm}\mbox{for $ \tau_1 D_1/(1+\lambda \tau_1)^2 \ll 1$} 
\end{eqnarray}
%This may be understood as follows. 
%Equation (27) shows that
%this approximation is valid under the condition of
%$\tau_1 D_1/(1+\lambda \tau_1)^2 \ll 1$.
Equation (27) implies that the approximation of
$\tilde{D}_1^{APP}$ is valid both
for (i) $\tau_1 \ll (1/\lambda, \: 1/D_1)$ 
and (ii) $\tau_1 \gg (1/\lambda, \: D_1/\lambda^2)$.
%this approximation is valid both for small and large $\tau_1$.

\subsection{Stationary distribution}

From the effective FPE given by Eq. (7), 
we get the stationary distribution $p(x)$ given by
\begin{eqnarray}
\ln p(x) & = & -\left(\frac{1}{2} \right) 
\ln[ \tilde{D}_0+ \tilde{D}_1 \:G(x)^2]+Z(x),
%&\propto & [\tilde{D}_0+\tilde{D}_1G_1^2(x)+\tilde{D}_2G_2^2(x)]^{-1/2}
%\;\exp[Z(x)]
\end{eqnarray}
with
\begin{eqnarray}
Z(x)=\int \frac{F(x)}
{[\tilde{D}_0+ \tilde{D}_1 G(x)^2]} \; dx. 
\end{eqnarray}
Because of the presence of multiplicative noise,
the stationary distribution generally has non-Gaussian
power-law structure \cite{Hasegawa07a}-\cite{Anten03}. 

In the white-noise limit ($\tau_m=0$), the stationary distribution
for the Langevin equation given by Eq. (1)
with $\eta_m=\sqrt{2 D_m}$ is expressed by Eqs. (28) and (29)
with $\tilde{D}_m=D_m$ in the Stratonovich representation.
Then the stationary distribution for colored noise
is expressed by
\begin{eqnarray}
p(x;D_0,D_1,\tau_0,\tau_1) &=&
p(x;\tilde{D}_0,\tilde{D}_1,0,0) 
\equiv p_{wn}(x;\tilde{D}_0,\tilde{D}_1),
\end{eqnarray}
where $p_{wn}(x;\tilde{D}_0,\tilde{D}_1)$ expresses
the stationary distribution for white noise.

In the case of $F(x)=-\lambda x+I$ and $G(x)=x$,
we get
\begin{eqnarray}
%\ln p(x) &=& 
%-\left( \frac{\lambda}{2 \tilde{D}_1}+\frac{1}{2} \right) 
%\ln (\tilde{D}_0+ \tilde{D}_1x^2)
%+ Y(x), \\
p(x) &\propto & (\tilde{D}_0+ \tilde{D}_1x^2)^{-(\lambda/2 \tilde{D}_1+1/2)} \exp[Y(x)],
\end{eqnarray}
with
\begin{eqnarray}
Y(x) = \frac{I}{\sqrt{\tilde{D}_0\tilde{D}_1} }. 
\tan^{-1} \left(\sqrt{ \frac{\tilde{D}_1}{\tilde{D}_0} } x \right),
\end{eqnarray}
\begin{eqnarray}
\tilde{D}_0 &=& \frac{D_0}{(1+\lambda \tau_0 ) }, \\
\tilde{D}_1 &=& \frac{D_1}{[1+(\tau_1I/\mu_s)]},
%\hspace{1.0cm}\mbox{($i= 1, 2$)} %\nonumber
\end{eqnarray}
where $\mu_s$ expresses the stationary value of $\mu$ [Eq. (23) ].

\subsection{Model calculations}
\subsubsection{Stationary properties}

In order to demonstrate the feasibility of our analytical theory,
we have performed model calculations.
Direct simulations (DSs) for Eqs. (1) and (2) have been performed 
by using the fourth-order Runge-Kutta method for period
of 1000 with a time mesh of 0.01.
Results of DSs are the average over hundred thousands trials
otherwise noticed. All quantities are dimensionless.

The $\tau_1$ dependence of
the ratio of $\tilde{D}_1/D_1$ is depicted in Fig.1, where
results calculated by the FIM and the approximation (APP)
given by Eq. (27) are shown by solid and dashed curves, respectively,
for $\lambda=1.0$, $D_0=0.01$, $D_1=0.2$ and $\tau_0=0.01$.
$\tilde{D}_1$ calculated by the APP is in good agreement
with that by the FIM, and the effective noise strength
is decreased with increasing $\tau_1$.
The difference between $\tilde{D}_1/D_1$ of
the FIM and APP, plotted by the chain curve,
is zero at $\tau_1 = 0$ with a maximum at $\tau_1 \sim 0.5$, and
decreased at larger $\tau_1 \:(> 1)$. 
The APP is fairly good for small $\tau_1$ and large $\tau_1$, 
as discussed after Eq. (27).

Figure 2 (a)-(f) show the stationary distribution $p(x)$
calculated by the FIM (solid curves), DS (dashed curves),
with the APP (chain curves)
and in the white-noise limit (WN: double-chain curves)
when $\tau_1$ is changed for fixed values of
$\lambda=1.0$, $D_0=0.01$, $D_1=0.2$ and $\tau_0=0.01$:
(a), (c) and (e) in normal scale, and (b), (d) and (f) in log scale.
Calculations show that
with increasing $\tau_1$, $p(x)$ becomes narrower, deviating
from results of WN.
Results of the FIM and APP are in fairly
good agreement with those of DS:
results of the APP is indistinguishable from those of the FIM.
Figures 2(b), (d) and (f) plotting $p(x)$ in log scale show
that results of the FIM and APP
are in fairly good agreement with that of DS
up to the order of $10^{-2}$ for $\tau_1=1.0$
and of $10^{-4}$ for $\tau_1=5.0$.

\subsubsection{Dynamical properties}

We have investigated the response to an applied pulse
input given by
\begin{equation}
I(t) = A \Theta(t-t_b)\:\Theta(t_e-t)+B,
\end{equation}
where $A=0.5$, $B=0.1$, $t_b=100$ and $t_e=200$, and
$\Theta(t)$ is the Heaviside function.
Time courses of $\mu(t)$ and $\gamma(t)$ are shown in Fig. 3(a)-(f),
where $\tau_1$ is changed for fixed values of
$\lambda=1.0$, $D_0=0.01$, $D_1=0.2$ and $\tau_0=0.01$.
With increasing $\tau_1$, $\mu(t)$ and $\gamma(t)$ induced
by an applied pulse at $t_b < t < t_e$ are decreased. 
This is because
they are given by
\begin{eqnarray}
\mu(t) &=& \frac{(A+B)}{(\lambda-\tilde{D}_1)}, \\
\gamma(t) &=& \frac{\tilde{D}_0}{(\lambda-2 \tilde{D}_1)}
+ \frac{(A+B)^2 \tilde{D}_1}{(\lambda-2 \tilde{D}_1)(\lambda-\tilde{D}_1)^2},
\hspace{1cm}\mbox{for $t_b < t < t_e$}
\end{eqnarray}
where $\tilde{D}_1$ is decreased with increasing $\tau_1$ as Fig. 1 shows.
The results of the FIM and APP are again in good agreement with
that of the DS: the FIM yields slightly better 
results than the APP as shown in Figs. 3(b) and 3(f).

Next we study the response to a sinusoidal input given by
\begin{equation}
I(t) = C \sin \omega t,
\end{equation}
where $C=0.5$, $\omega=2 \pi/T_p$ and $T_p=100$.
Figures 4(a), (c) and (f) show time courses of 
$\mu(t)$, and Figs. 4(b), (d) and (f) those of $\gamma(t)$
when $\tau_1$ is changed for fixed values of
$\lambda=1.0$, $D_0=0.01$, $D_1=0.2$ and $\tau_0=0.01$.
With increasing $\tau_1$, the magnitude of $\mu(t)$ 
is decreased. This is understood  
from an analysis with the use of Eq. (19), which yields
\begin{equation}
\mu(t) = \frac{C}{\sqrt{(\lambda-\tilde{D}_1)^2+\omega^2}}
\: \sin(\omega t - \phi),
\end{equation}
with
\begin{equation}
\phi=\tan^{-1}\left( \frac{\omega}{\lambda-\tilde{D}_1} \right).
\end{equation}
Equation (39) shows that with increasing $\omega$ 
({\it i.e.} decreasing $T_p$), the magnitude of $\mu(t)$
is decreased, representing a character of the low-pass filter.

%\newpage
\section{Langevin model subjected to one additive and
two multiplicative colored noise}

%\section{Case of one additive and
%two independent multiplicative colored noise}

\subsection{Effective Langevin equation}

We have assumed the Langevin model subjected to one additive
($\eta_0$) and two multiplicative 
colored noise ($\eta_1$, $\eta_2$), as given by
\begin{eqnarray}
\frac{d x}{dt}\!\!&=&\!\!F(x) + \eta_0(t)+ G_1(x) \eta_1(t)
+ G_2(x) \eta_2(t), 
\end{eqnarray}
with
\begin{eqnarray}
\frac{d \eta_{m}(t)}{d t}
=-\frac{1}{ \tau_{m} }\eta_{m}
+ \frac{ \sqrt{2 D_{m}} }{\tau_{m}} \xi_{m}(t),
\hspace{1.0cm}\mbox{($m=0,1,2$)} %\nonumber
\end{eqnarray}
where $F(x)$, $G_1(x)$ and $G_2(x)$ express arbitrary functions 
of $x$, and $\xi_m$ are independent zero-mean white noise
with correaltion:
\begin{eqnarray}
\langle \xi_{m}(t)\:\xi_{n}(t') \rangle
&=& \delta_{mn}\delta(t-t').
%\hspace{1cm}\mbox{($i=0,1,2$)}
\end{eqnarray}

Applying the FIM \cite{Liang04} to the model under consideration,
we get the effective FPE given by
(details being given in the Appendix):
\begin{eqnarray}
\frac{\partial}{\partial t}p(x,t) 
&=& - \frac{\partial }{\partial x} \tilde{F}(x)p(x,t)
+ \tilde{D}_0 \frac{\partial^2 }{\partial x^2} p(x,t) \nonumber \\
&+& \tilde{D}_1 \frac{\partial }{\partial x}G_1(x)
\frac{\partial }{\partial x}G_1(x) p(x,t)
+ \tilde{D}_2 \frac{\partial }{\partial x}G_2(x)
\frac{\partial }{\partial x}G_2(x) p(x,t),
\end{eqnarray}
from which we get the effective Langevin equation:
\begin{eqnarray}
\frac{d x}{dt}\!\!&=&\!\!F(x) + \sqrt{2\tilde{D}_0}\: \xi_0(t)
+ \sqrt{2 \tilde{D}_1} \:G_1(x) \:\xi_1(t)
+ \sqrt{2 \tilde{D}_2} \:G_2(x) \:\xi_2(t),
\end{eqnarray}
with
\begin{eqnarray}
\tilde{D}_0 &=& \frac{D_0}{(1-\tau_0  \langle F' \rangle)}, \\
\tilde{D}_{m} &=& \frac{D_{m}}
{[1-\tau_{m} (\langle F' \rangle - \langle  F G'_m/G_m \rangle)]},
\hspace{1.0cm}\mbox{($m=1,2$)} %\nonumber
\end{eqnarray}
where $F'=dF/dx$ and $G'_m=dG_m/dx$, and the bracket 
$\langle \cdot \rangle$ stands for the average over $p(x,t)$: 
\begin{eqnarray}
\langle Q(x) \rangle &=& \int Q(x) \: p(x,t)\: dx.
\end{eqnarray}

\subsection{Equations of motion for mean and variance}

By using the effective FPE given by Eq. (45), we can obtain
equations of motion for mean ($\mu$) and variance ($\gamma$) 
defined by
\begin{eqnarray}
\mu &=& \langle x \rangle, \\
\gamma &=& \langle x^2 \rangle - \langle x \rangle^2.
\end{eqnarray}
When $F(x)$ and $G_m(x)$ are given by
\begin{eqnarray}
F(x) &=&- \lambda x+ I, \\ 
G_m &=& a_m (x-e_m),
\hspace{1cm}\mbox{($m=1,2$)}
\end{eqnarray}
where $\lambda$ is the relaxation rate, $I$ an input, and
$a_m$ and $e_m$ constants, 
we get equations of motion for $\mu$ and $\gamma$ given by 
\cite{Hasegawa07a}
\begin{eqnarray}
\frac{d \mu}{d t} &=& -\lambda \mu + I
+ \tilde{D}_1(\mu-e_1) +\tilde{D}_2 (\mu-e_2), \\
\frac{d \gamma}{dt} &=& - 2 \lambda \gamma 
+ 4 (\tilde{D}_1+\tilde{D}_2 ) \gamma 
+ 2 \tilde{D}_2 (\mu-e_1)^2
+ 2 \tilde{D}_2 (\mu-e_2)^2 + 2 \tilde{D}_0,
\end{eqnarray}
with
\begin{eqnarray}
\tilde{D}_0 &=& \frac{D_0}{(1+ \lambda \tau_0)}, \\
\tilde{D}_{m} &=& \frac{D_{m}}
{[1+\tau_{m} (-\lambda e_m+I)/(\mu-e_m)]}. 
\hspace{1cm}\mbox{(for $m=1,2$)} 
\end{eqnarray}
It is necessary to self-consistently solve Eqs. (53)-(56)
for $\mu$, $\gamma$,
$\tilde{D}_{0}$, $\tilde{D}_{1}$ and $\tilde{D}_{2}$.

Stationary values of $\mu$ and $\gamma$ are implicitly given by
\begin{eqnarray}
\mu_s &=& \frac{(I-\tilde{D}_1 e_1-\tilde{D}_2 e_2)}
{(\lambda - \tilde{D}_1-\tilde{D}_2)}, \\
\gamma_s &=& \frac{[\tilde{D}_0+ \tilde{D}_1(\mu_s-e_1)^2
+\tilde{D}_2(\mu_s-e_2)^2]}
{[\lambda - 2(\tilde{D}_1+ \tilde{D}_2)]},
\end{eqnarray}
with $\tilde{D}_m$ ($m=0,1,2$) given by Eqs. (55) and (56)
with $\mu=\mu_s$.
Equations (57) and (58) show that $\mu_s$ and $\gamma_s$
diverge for $(\tilde{D}_1 + \tilde{D}_2) > \lambda$ 
and $(\tilde{D}_1+\tilde{D}_2) > \lambda/2$, respectively.
Equation (27) suggests that the approximation given by
\begin{eqnarray}
\tilde{D}_{m} & \simeq & \frac{D_{m}}
{(1+\lambda \tau_{m}) } \equiv \tilde{D}_m^{APP},
\hspace{1cm}
\mbox{for $\tau_m D_m/(1+\lambda \tau_m)^2 \ll 1$ ($m=1,2$)}
\end{eqnarray}
may be valid both for small $\tau_m$ and large $\tau_m$,
as will be numerically shown in Fig. 5
\cite{Note1}. 

\subsection{Stationary distribution}

From the effective FPE of Eq. (45), we get
the stationary distribution $p(x)$ given by
\cite{Hasegawa07a}
\begin{eqnarray}
\ln p(x) &= & -\left(\frac{1}{2} \right) 
\ln[\tilde{D}_0 +\tilde{D}_1 G_1^2(x)
+\tilde{D}_2 G_2^2(x)]+Z(x),
\end{eqnarray}
with
\begin{eqnarray}
Z(x)&=&\int \frac{F(x)}
{[\tilde{D}_0+\tilde{D}_1 G_1^2(x)+\tilde{D}_2 G_2^2(x)]} \; dx. 
\end{eqnarray}

In the white-noise limit ($\tau_{m}=0$),
the stationary distribution of the Langevin model 
given by Eq. (41)
with $\eta_{m}= \sqrt{2 D_{m}}$ ($m=0,1,2$),
is expressed by 
\cite{Hasegawa07a}-\cite{Anten03}
\begin{eqnarray}
\ln p(x) & = & Z(x) -\left(\frac{1}{2} \right) 
\ln[D_0 +D_1 G_1^2(x)+D_1 G_1^2(x)],
\end{eqnarray}
with
\begin{eqnarray}
Z(x)=\int \frac{F(x)}
{[D_0 +D_1 G_1^2(x)+D_2 G_2^2(x)]} \; dx, 
\end{eqnarray}
in the Stratonovich representation. 
It agrees with the distribution given by Eqs. (60) and (61) with
$\tilde{D}_m=D_m$.

When $F(x)$ and $G_m(x)$ are given by Eqs. (51) and (52),
we get
\begin{eqnarray}
%\ln p(v) 
%&=& -\left(\frac{\lambda}{2d_2}+\frac{1}{2} \right)
%\ln (d_2 v^2+d_1 v + d_0) \nonumber \\
%&+& \left(\frac{2 I d_2+\lambda d_1}{d_2 \sqrt{4 d_0 d_2-d_1^2}} \right)
%\tan^{-1}\left(\frac{2 d_2 v+d_1}{\sqrt{4 d_0 d_2-d_1^2}} \right) 
p(v) &\propto&  (d_2 v^2+d_1 v + d_0)^{-(\lambda/2d_2+1/2)}
\;\exp[Y(v)],
\end{eqnarray}
with
\begin{eqnarray}
Y(v) &=& 
\left(\frac{2 I d_2+\lambda d_1}{d_2 \sqrt{4 d_0 d_2-d_1^2}} \right)
\tan^{-1}\left(\frac{2 d_2 v+d_1}{\sqrt{4 d_0 d_2-d_1^2}} \right),
\end{eqnarray}
where
\begin{eqnarray}
\tilde{D}_0 &=& \frac{D_0}{(1+ \lambda \tau_0)}, \\
\tilde{D}_{m} &=& \frac{D_{m}}
{[1+\tau_{m} (-\lambda e_m+I)/(\mu_s-e_m)]}, 
\hspace{1cm}\mbox{(for $m=1,2$)} 
\end{eqnarray}
\begin{eqnarray}
d_0 &=& \tilde{D}_0 + \tilde{D}_1 a_1^2 e_1^2
+ \tilde{D}_2 a_2^2 e_2^2 \\
d_1 &=& -2( \tilde{D}_1 a_1^2 e_1+\tilde{D}_2 a_2^2 e_2)  \\
d_2 &=&  \tilde{D}_1 a_1^2 + \tilde{D}_1 a_1^2,
\end{eqnarray}
$\mu_s$ denoting the stationary value [Eqs. (57) and (58)].

It is easy to see that for additive noise only 
($D_1=D_2=0$), the distribution becomes the Gaussian
given by
\begin{eqnarray}
p(x) \propto \exp\left[-\frac{\lambda}{2 d_0}
\left(x-\frac{I}{\lambda} \right)^2 \right].
\end{eqnarray}
When multiplicative noise is included, $p(x)$
becomes the non-Gaussian distribution with power-law tails.

\subsection{Model calculations}
\subsubsection{Stationary properties}

Figure 5 shows $\tilde{D}_m/D_m$ ($m=1,2$) as a function of
$\tau_1$ calculated by the FIM for $m=1$ (solid curve)
and $m=2$ (dashed curve) with fixed values
of $\tau_2/\tau_1=10$,
$\lambda=1.0$, $D_0=0.01$, $D_1=0.1$, $D_2=0.2$ and $\tau_0=0.01$.
With increasing $\tau_1$ ($=\tau_2/10$),
$\tilde{D}_1/D_1$ and $\tilde{D}_2/D_2$ are gradually decreased.
In order to examine the validity of 
the approximation (APP) given by $\tilde{D}_m^{APP}$ in Eq. (59),
we show differences between $\tilde{D}_m/D_m$ of the FIM and APP
for $m=1$ (chain curve) and $m=2$ (double-chain curve).
Both the differences start from zero at $\tau_1=0$,
have maxima at $\tau_1 \sim 0.5-1$, and
are decreased at larger $\tau_1$. This shows that
the APP is valid both for small $\tau_1$ and large $\tau_1$,
whose behavior is similar to that shown in Fig. 1.

Figures 6(a)-(f) show the stationary distribution $p(x)$
calculated by the FIM (solid curves), DS (dashed curves),
the APP (chain curves)
and in the white-noise limit (double chain curves),
when $\tau_1$ and $\tau_2$ are changed with
a fixed ratio of $\tau_2/\tau_1=10$, and
$\tau_0=0.01$, $I=0.5$, $\lambda=1.0$,
$D_0=0.01$, $D_1=0.1$ and $D_2=0.2$:
Fig. 6(a), (c) and (e) are plotted in normal scale
while Fig. 6(b), (d) and (f) in log scale.
With increasing the relaxation time for multiplicative colored
noise, the width of $p(x)$ is decreased and its profile approaches
the Gaussian.

\subsubsection{Dynamical properties}

Responses of $\mu(t)$ to a pulse input given by Eq. (35)
are shown in Figs. 7(a), (c) and (e) by changing 
the relaxation time with a fixed value of $\tau_2/\tau_1=10$.
They are calculated by the FIM (solid curves), DS (dashed curves),
the APP (chain curves) 
and in the white-noise limit (WN: double-chain curves).
Similarly the response of $\gamma(t)$ are plotted
in Fig. 7(b), (d) and (f).
With increasing the relaxation time of colored noise, 
the effective noise strength
is decreased, and then the values of $\mu(t)$ and $\gamma(t)$
at $100 < t < 200$ is decreased, as shown by Eqs. (39) and (40).  
Figures 7(d) and 7(f) show that the FIM yield slightly
better results than the APP.
The general trend of the effect of the relaxation time
on $p(x)$, $\mu(t)$ and $\gamma(t)$ shown in Figs. 6 and 7
is the same as that shown in Figs. 2 and 3.

%\newpage

\section{Discussion}
\subsection{A controversy on the subthreshold voltage distribution}
In recent years, a controversy has been made on 
the subthreshold voltage distribution of a leaky 
integrate-and-fire model \cite{Rudolph03}-\cite{Rudolph06}. 
The adopted model includes conductance-based stochastic
excitatory and inhibitory synapses as well as noisy inputs, 
as given by \cite{Rudolph03}-\cite{Rudolph06}
\begin{eqnarray}
C \frac{d v}{d t} &=& - g_L(v-E_L) -\frac{1}{a}I_{syn}(t),
\end{eqnarray}
with
\begin{eqnarray}
I_{syn}(t) &=& g_e(v-E_e)+g_i(v-E_i) - I_I(t).
\end{eqnarray}
Here $C$ denotes the membrane capacitance,
$a$ the membrane area, 
$g_L$ and $E_L$ the leak conductance and reversal potential, and
$g_{e,i}$ and $E_{e,i}$ are the noisy conductances
and reversal potentials of excitatory ($e$) and inhibitory $i$
synapses, respectively. Stochastic $g_{e,i}$ and
noisy additional input $I_I(t)$ are assumed to be described
by the OU process given by
\begin{eqnarray}
\frac{d g_e}{d t} &=& - \frac{1}{\tau_e}(g_e-g_{e0})
+\sqrt{\frac{2 \sigma_e^2}{\tau_e}} \xi_e(t), \\
\frac{d g_i}{d t} &=& - \frac{1}{\tau_i}(g_i-g_{i0})
+\sqrt{\frac{2 \sigma_i^2}{\tau_i}} \xi_i(t), \\
\frac{d I_I}{d t} &=& - \frac{1}{\tau_I}(I_I-I_{0})
+\sqrt{\frac{2 \sigma_I^2}{\tau_I}} \xi_I(t), 
\end{eqnarray}
where $\xi_e$ {\it et el.} are independent zero-mean white noise
with correlation:
\begin{eqnarray}
\langle \xi_{\kappa}(t)\:\xi_{\kappa'}(t') \rangle
&=& \delta_{\kappa \kappa'}\delta(t-t').
\hspace{1cm}\mbox{($\kappa=I,e,i$)}
\end{eqnarray}
In the first paper of Rudolph and Destexhe (RD1) \cite{Rudolph03}, 
they derived an expression for the stationary distribution function
of the system described by Eqs. (72)-(76).
In their second paper (RD2) \cite{Rudolph05}, they modified 
their expression to cover  a larger parameter regime. 
Lindner and Longtin \cite{Lindner06} criticized that the result of
RD1 does not reconcile the result of white-noise limit
and that the extended expression of RD2 does not solve the
colored-noise problem though it is much better than that of RD1.
In the third paper of Rudolph and Destexhe (RD3) \cite{Rudolph06},
they claimed that the result of RD2 is the best from a comparison 
among various approximate analytic expressions for the
stationary distribution. 
It has been controversy which of approximate analytic 
expressions having been proposed so far may best explain the result
of DSs.

It is worthwhile to apply our method mentioned in Sec. 3
to the system given by Eq. (72)-(76), which are rewritten as
\begin{eqnarray}
\frac{dv}{dt}\!\!&=&\!\!F(v) + G_I \eta_I(t)+ G_e(v) \eta_e(t)
+ G_i(v) \eta_i(t), 
\end{eqnarray}
with
\begin{eqnarray}
\frac{\partial \eta_{\kappa}(t)}{\partial t}
=-\frac{1}{ \tau_{\kappa} }\eta_{\kappa}
+ \frac{ \sqrt{2 D_{\kappa}} }{\tau_{\kappa}} \xi_{\kappa}(t),
\hspace{1.0cm}\mbox{($\kappa=e, i, I$)} %\nonumber
\end{eqnarray}
where
\begin{eqnarray}
F(v) &=& -\frac{1}{C}g_L(v-E_m) 
-\frac{1}{C a}[g_{e0}(v-E_e)
+g_{i0}(v-E_i)-I_{0}], \\
G_I &=& \frac{1}{Ca}, \\
G_{e,i}(v) &=& -\frac{1}{Ca}(v-E_{e,i}), \\
D_{I,e,i} &=& \tau_{I,e,i} \sigma_{I,e,i}^2.
\end{eqnarray}

By using Eqs. (64)-(70), we get
the stationary distribution $p(v)$ given by
\begin{eqnarray}
p(v) &\propto&  (d_2 v^2+d_1 v + d_0)^{-(\lambda/2d_2+1/2)}
\;\exp[Y(v)],
\end{eqnarray}
with
\begin{eqnarray}
Y(v) &=& 
\left(\frac{2 c_0 d_2+\lambda d_1}{d_2 \sqrt{4 d_0 d_2-d_1^2}} \right)
\tan^{-1}\left(\frac{2 d_2 v+d_1}{\sqrt{4 d_0 d_2-d_1^2}} \right),
\end{eqnarray}
where
\begin{eqnarray}
\lambda &=& \frac{1}{C}g_L +\frac{1}{Ca} (g_{e0}+ g_{i0}), \\
c_0 &=& \frac{1}{C}g_L E_m 
+ \frac{1}{Ca}(g_{e0}E_e + g_{i0} E_i + I_{I0}), \\
d_0 &=& \frac{1}{C^2 a^2}
(\tilde{D}_I + \tilde{D}_e E_e^2 + \tilde{D}_i E_i^2), \\
d_1 &=& - \frac{2}{C^2 a^2}
(\tilde{D}_e E_e + \tilde{D}_i E_i), \\
d_2 &=& \frac{1}{C^2 a^2}(\tilde{D}_e+\tilde{D}_i).
\end{eqnarray}
With the use of Eqs. (57) and (58), 
$\tilde{D}_{\kappa}$ in Eqs. (88)-(90) 
are expressed by
\begin{eqnarray}
\tilde{D}_I &=& \frac{D_I}{(1+ \lambda \tau_I)}, \\
\tilde{D}_{e} &=& \frac{D_{e}}
{[1+\tau_{e} (-\lambda E_e+c_0)/(\mu_s-E_e)]}, \\
\tilde{D}_{i} &=& \frac{D_{i}}
{[1+\tau_{i} (-\lambda E_i+c_0)/(\mu_s-E_i)]}.
\end{eqnarray}
Here $\mu_s$ denotes the stationary value of $v$ 
which is determined by the self-consistent equations
for $\mu_s$, $\gamma_s$, $\tilde{D}_{e}$ and $\tilde{D}_{i}$
[as Eqs. (57) and (58)], though their explicit expressions 
are not necessary for our discussion. 

In the limit of small relaxation times,
Eqs. (92) and (93) yield
\begin{eqnarray}
\tilde{D}_{e} & \simeq & \frac{D_{e}}
{( 1+\lambda \tau_{e}) }, 
\hspace{1cm}\mbox{(for $\tau_e D_e \ll 1$)}
\\
\tilde{D}_{i} & \simeq & \frac{D_{i}}
{(1+\lambda \tau_{i})}. 
\hspace{1cm}\mbox{(for $\tau_i D_i \ll 1$)}
\end{eqnarray}
This is nothing but the approximation introduced
in RD2 \cite{Rudolph05}.

In the white-noise limit ($\tau_{I,e,i}=0$),
the stationary distribution of the model given by Eq. (78)
with $\eta_{\kappa}= \sqrt{2 D_{\kappa}}$ ($\kappa=I,e,i$)
is given by 
\cite{Hasegawa07a}-\cite{Anten03}
\begin{eqnarray}
\ln p_{wn}(v;D_I,D_e,D_i) & = & Z(v) -\left(\frac{1}{2} \right) 
\ln[D_I G_I^2+D_e G_e^2(v)+D_i G_i^2(v)],
\end{eqnarray}
with
\begin{eqnarray}
Z(v)=\int \frac{F(v)}
{[D_I G_I^2 +D_e G_e^2(v)+D_i G_i^2(v)]} \; dv, 
\end{eqnarray}
in the Stratonovich representation. 
We note that our stationary distribution given by Eqs. (84)-(90)
is consistent in the white-noise limit, and that the distribution
for colored noise, is expressed by
\begin{eqnarray}
p(v;D_I,D_e,D_i,\tau_I,\tau_e,\tau_i)
&=& p_{wn}(v;\tilde{D}_I,\tilde{D}_e,\tilde{D}_i),
\end{eqnarray}
where $\tilde{D}_I$, $\tilde{D}_e$ and $\tilde{D}_i$ 
given by Eqs. (91)-(93) take account of
effects of relaxation times.
If we adopt approximate expressions for 
$\tilde{D}_e$ and $\tilde{D}_i$ given by Eqs. (94) and (95),
we get
\begin{eqnarray}
p(v;D_I,D_e,D_i,\tau_I,\tau_e,\tau_i)
&=& p_{wn}\left(v;\frac{D_I}{(1+\lambda \tau_I)},
\frac{D_e}{(1+\lambda \tau_e)},\frac{D_i}{(1+\lambda \tau_i)} \right).
\nonumber \\
&& \hspace{3cm}\mbox{(for $\tau_e D_e \ll 1$ and $\tau_i D_i \ll 1$)}
\end{eqnarray}

Lindner and Longtin have pointed out that
the first solution of RD1 is given by \cite{Lindner06}
\begin{eqnarray}
p_{RD1}(v;D_I,D_e,D_i,\tau_I,\tau_e,\tau_i)
&=& p_{wn}\left(v;\frac{D_I}{2},\frac{D_e}{2},\frac{D_i}{2} \right).
\end{eqnarray}
With the use of the Fourier transform of stochastic equations,
Rudolph and Destexhe have obtained in RD2 \cite{Rudolph05}, 
the stationary distribution with $D_I=0$ given by 
\begin{eqnarray}
p_{RD2}(v;D_I=0,D_e,D_i,\tau_I,\tau_e,\tau_i)
&=& p_{wn}\left(v;D_I=0,\frac{D_e}{(1+\lambda \tau_e)},
\frac{D_i}{(1+\lambda \tau_i)} \right).
\end{eqnarray}
It is noted that Eq. (101) coincides with Eq. (99) for $D_I=0$.
Model calculations in Sec. 3.4 have shown that the approximation
given by Eq. (59) yields a good result. This is true
also for the approximation given by Eqs. (94) and (95).
This explains to some extent
the reason why the approximation adopted in
RD2 provides us with good results,
as claimed in RD3 \cite{Rudolph06}.

\subsection{A comparison with previous approximations}

We have discussed, in Sec. 2, the dynamics of the Langevin model
subjected to colored model, by using the FIM.
It is interesting to make a comparison among 
several approximate, analytical methods
having been proposed so far
for the Langevin model subjected to colored noise:
%We consider the model subjected to multiplicative colored noise:
\begin{eqnarray}
\frac{dx}{dt}\!\!&=&\!\!F(x) +  G(x) \eta_1(t), 
\end{eqnarray}
where $\eta_1(t)$ is described by the OU process of Eq. (2).

First we apply the UCNA to Eq. (102) \cite{Hanggi95,Jung87}.
Taking the derivative of Eq.(102) with respect to $t$,
eliminating the variable $\eta_1$ with the use of
Eq. (2), and neglecting the $\ddot{x}$ term after the UCNA, we get
the effective Langevin equation given by
\begin{eqnarray}
\frac{dx}{dt}\!\!&=&\!\! \tilde{F}(x) +  \tilde{D}_1 G(x) \xi_1(t), 
\end{eqnarray}
with
\begin{eqnarray}
\tilde{F}^{U} &=& \frac{F(x)}{[1-\tau_1 (F'-F G'/G)]}, \\
\tilde{D}_1^{U} &=& \frac{D_1}{[1-\tau_1 (F'-F G'/G)]^2}.
\end{eqnarray}
The stationary distribution is given by
\begin{eqnarray}
p(x) &\propto & \exp[-\ln G(x)+Y(x)],
\end{eqnarray}
with
\begin{eqnarray}
Y(x) &=& \left( \frac{1}{\tilde{D}_1} \right)
\int \frac{\:\tilde{F}(x)}{G(x)^2} \;dx.
\end{eqnarray}
In the case of $F(x)=-\lambda x$ and $G(x)=x$, we get
\begin{eqnarray}
p(x) &\propto& x^{-[\lambda(1+\lambda \tau_1)/D_1+1]},
\end{eqnarray}
which agrees with Eqs. (31) and (32) with $I=Y=0$.

Similarly,
when the system is subjected to additive colored noise only
\begin{eqnarray}
\frac{dx}{dt}\!\!&=&\!\!F(x) +  \eta_0(t), 
\end{eqnarray}
the UCNA yields the effective Langevin equation given by
\begin{eqnarray}
\frac{dx}{dt}\!\!&=&\!\! \tilde{F}(x) +  \tilde{D}_0 \xi_0(t), 
\end{eqnarray}
with
\begin{eqnarray}
\tilde{F}^{U} &=& \frac{F(x)}{(1-\tau_0 F')}, \\
\tilde{D}_0^{U} &=& \frac{D_0}{(1-\tau_0 F')^2}.
\end{eqnarray}
The stationary distribution is given by
\begin{eqnarray}
p(x) &\propto & \exp[Z(x)],
\end{eqnarray}
with
\begin{eqnarray}
Z(x) &=& \left(\frac{1-\tau_0 F'}{D_0} \right) \int\:F(x) \;dx.
\end{eqnarray}

When a system is subjected to {\it multiplicative} 
colored noise only, as given by Eq. (102), 
we may employ the method of a change 
of variable, with which it is transformed to
that subjected to {\it additive} noise \cite{Risken96}:
\begin{eqnarray}
\frac{d y}{d t} &=& \tilde{F}(y) + \eta_1(t),
\end{eqnarray}
where
\begin{eqnarray}
y &=& \int \frac{dx}{G(x)} \equiv K(x), \\
\tilde{F}(y) &=& \frac{F(K^{-1}(y))}{G(K^{-1}(y))}.
\end{eqnarray}
We get the effective Langevin equation given by Eq. (103) with
\begin{eqnarray}
\tilde{F}^{C} &=& F(x), \\
\tilde{D}_1^{C} &=& \frac{D_1}{[1-\tau_1 (F'-F G'/G)]}. 
\end{eqnarray}
In the case of $F(x)=-\lambda x$ and $G_1(x)=x$, we get
the distribution of $p(x)$ given by
\begin{eqnarray}
p(x) &\propto& x^{-[\lambda(1+\lambda \tau_1)/D_1+1]},
\end{eqnarray}
which agrees with Eqs. (108).

Table 1 summarizes a comparison among various approximate
methods.
It is noted that although 
the effective Langevin equation
is rather different depending on the methods,
the stationary distribution given by Eq. (108) or (120)
agrees each other in the linear Langevin model
for which the ratio of $\tilde{F}(x)/\tilde{D}_1(x)$ is 
the same [Eq. (107)].
Difference among the methods may be, however, realized 
in dynamical properties.
Figure 8(a) and 8(b) show responses of $\mu(t)$ and $\gamma(t)$,
respectively, to an applied pulse input given by Eq. (35),
calculated by the FIM (solid curves), DS (dashed curves)
and the UCNA (chain curves)
with $\lambda=1.0$, $D_1=0.2$ and $\tau_1=1.0$. 
We note that $\mu(t)$ and $\gamma(t)$ at $100 < t < 200$ 
calculated by the FIM is in good agreement with those of DS.
Those of the UCNA are, however, much larger than those of DS.
This is easily understood by Eqs. (36), (37), (104) and (105), 
from which we get (for $D_0 \ll D_1$)
\begin{eqnarray}
\mu^{F}(t) &\simeq& \frac{(A+B)}{[\lambda-D_1/(1+\lambda \tau_1)]}, 
\hspace{5cm}\mbox{(for FIM)} \\
\mu^{U}(t) &\simeq& (1+\lambda \tau_1)\: \mu^{F}(t),
%&=& \frac{A+B}{[\lambda/(1+\lambda \tau_1)-D_1/(1+\lambda \tau_1)^2]}, 
\hspace{6cm}\mbox{(for UCNA)} \\
\gamma^{F}(t) &\simeq & \frac{(A+B)^2 D_1/(1+\lambda \tau_1)}
{[\lambda-2 D_1/(1+\lambda \tau_1)]
[\lambda-D_1/(1+\lambda \tau_1)]^2},
\hspace{1cm}\mbox{(for FIM)} \\
\gamma^{U}(t) &\simeq & (1+\lambda \tau_1)^2\:\gamma^{F}(t). 
%&\simeq & \frac{(A+B)^2 D_1/(1+\lambda \tau_1)^2}
%{[\lambda/(1+\lambda \tau_1)-2 D_1/(1+\lambda \tau_1)^2]
%[\lambda/(1+\lambda \tau_1)-D_1/(1+\lambda \tau_1)^2]^2}. \nonumber \\
\hspace{6cm}\mbox{(for UCNA)}
\end{eqnarray}
Similar results are obtained also for sinusoidal inputs:
responses of $\mu(t)$ and $\gamma(t)$
to a sinusoidal input given by Eq. (38) are shown in
Figs. 8(c) and 8(d), respectively. We note 
that their magnitudes calculated by the UCNA
are again much larger than those of DS and the FIM.

It has been claimed that the UCNA is justified by 
the FIM \cite{Colet89,Wio89,Fuentes02}.
Although the FIM starts from the formally
exact expression for the probability distribution, 
an actual evaluation
has to adopt some kinds of approximations such as $\ddot{x}=0$ and
$(\dot{x})^n =0$ for $n \geq 2$ just as in the UCNA
\cite{Hanggi95,Jung87}.
The result of Ref. \cite{Colet89,Wio89,Fuentes02} obtained by the FIM
is different from that of Ref. \cite{Sancho82,Liang04,Wu07}
derived by the alternative FIM: the final result using the FIM 
depends on the adopted approximations.

\section{Conclusion}

We have extended the FIM approach such that we may discuss
dynamics of the Langevin model subjected to
additive and/or multiplicative colored noises, 
combined with equations of motion for mean and variance of 
a state variable $x$.
The stationary probability distribution and the dynamical 
response to time-dependent inputs have been discussed
for two cases of colored noise:
(a) one additive and one multiplicative colored noise, 
and (b) one additive and two multiplicative colored noise. 
Our conclusions are summarized as follows:

\noindent
(i) calculated results for the both cases (a) and (b) of the FIM 
are in good agreement with those of DS
for not only stationary but also dynamical properties: the latter 
can not be well accounted for by the existing, approximate 
analytical methods like the UCNA \cite{Jung87},

\noindent
(ii) with increasing the relaxation time of colored noise,
the width of the stationary distribution
$p(x)$ becomes narrower and 
its non-Gaussian form approaches the Gaussian because
the effective noise strength of $\tilde{D}_m$ becomes smaller than the
original noise strength of $D_m$, and

%\noindent
%(2) with increasing the relaxation time
%of colored noise, magnitudes of $\mu(t)$ and $\gamma(t)$ induced 
%by applied inputs are decreased,

\noindent
(iii) the approximations given by Eqs. (27) and (59) 
for the cases (a) and (b), respectively,
%(or Eqs. (94) and (95) obtained in RD2 \cite{Rudolph05})
derived by the FIM are valid for both small and
large relaxation times, which supports
the result of RD2 and RD3 \cite{Rudolph05,Rudolph06}.

\noindent
%Although the FIM provides us with a fairly nice 
%qualitative description of properties 
%of stochastic equations including colored noise,
%it is necessary to pursue a procedure 
%to obtain a better or exact result.  
The item (i) implies that the present FIM approach may be
applicable to a wide class of realistic models 
for physical systems subjected to noise sources 
with finite correlation time.
Recently we have proposed a generalized Langevin-type rate-code
neuronal model including multiplicative white noise 
\cite{Hasegawa07c,Hasegawa07d}. 
It would be interesting to study effects of finite correlation time
of colored noise
on firing rates in neuronal ensembles
based on the rate-code hypothesis, which is an alternative
to the temporary-code hypothesis \cite{Rieke96,Gerstner02}.

\section*{Acknowledgements}
This work is partly supported by
a Grant-in-Aid for Scientific Research from the Japanese 
Ministry of Education, Culture, Sports, Science and Technology.  

%\newpage
\vspace{1cm}
\appendix

\noindent
{\large\bf Appendix: Derivation of the effective Fokker-Planck equation}
\renewcommand{\theequation}{A\arabic{equation}}
\setcounter{equation}{0}

By applying the functional-integral method 
to the Langevin model given by
\begin{eqnarray}
\frac{d x}{dt}\!\!&=&\!\!F(x) + \sum_m G_m(x) \eta_m(t), 
\end{eqnarray}
with
\begin{eqnarray}
\frac{d \eta_{m}(t)}{d t}
&=& -\frac{1}{ \tau_{m} }\eta_{m}
+ \frac{ \sqrt{2 D_{m}} }{\tau_{m}} \xi_{m}(t), 
\end{eqnarray}
\begin{eqnarray}
\langle \xi_m(t) \xi_n(t')  \rangle
&=& \delta_{mn} \delta(t-t'),
\end{eqnarray} 
we may obtain the expression for the probability distribution:
$ p(x,t)= \langle\delta(x(t)-x) \rangle $ given by
\cite{Sancho82,Liang04}
\begin{eqnarray}
\frac{\partial}{\partial t} p(x,t) &=&
- \frac{\partial}{\partial x} F(x) p(x,t)
- \sum_{m} \frac{\partial}{\partial x} G_m(x) 
\langle \eta_m(t) \delta(x(t)-x) \rangle,
\end{eqnarray}
where the bracket 
$ \langle \cdot \rangle $ denotes the average over $p(x,t)$.
Although Eq. (A4) is a formally exact expression,
it is evitable to employ some approximations to perform
actual calculations.
In order to evaluate the average
$\langle \eta_m(t) \delta(x(t)-x) \rangle $, we use
the Novikov theorem \cite{Novikov65}:
\begin{eqnarray}
\langle \eta_m(t) \Phi(\eta_1, \eta_2) \rangle
&=&\int_0^t \:dt' \:\sum_n c_{mn}(t,t')  
\left< \frac{\delta \Phi(\eta_1, \eta_2)}{\delta \eta_n} \right>,
\end{eqnarray}
where $\Phi(\eta_1, \eta_2)$ denotes a function of
$\eta_1$ and $\eta_2$, and $c_{mn}$ is their correlation function
given by Eq. (5).
From Eqs. (5) and (A5), we get
\begin{eqnarray}
\langle \eta_m(t) \delta(x(t)-x) \rangle
&=& - \int_0^t \:dt' \:c_{mm}(t,t')
\left< \frac{\delta( \delta(x(t)-x)) }{\delta x} 
\frac{\delta x}{\delta \eta_m(t')} \right>, 
\\
&=& - \frac{\partial }{\partial x} \int_0^t \;dt' \:c_{mm}(t,t')
\left< \delta(x(t)-x) 
\frac{\delta x}{\delta \eta_m(t')} \right>.
\end{eqnarray}
After integrating Eq. (1), we get
\begin{eqnarray}
x(t)=x(0)+\int_0^t \:ds \:[F(x(s))+ \sum_m G_m(x(s)) \eta_m(s)].
\end{eqnarray}
The functional derivative of $x(t)$ of Eq. (A8) with respect to
$\eta_m(t')$ becomes
\begin{eqnarray}
\frac{\delta x(t)}{\delta \eta_m(t')}
&=& G_m(x(t'))+ \int_{t'}^{t}\:ds \: [F'(x(s))
+ \sum_n G'_n(x(s)) \eta_m(s)] \frac{\delta x(s)}{\delta \eta_m(t')}.
\end{eqnarray}
The derivative of Eq. (A9) with respect to $t$ is given by
\begin{eqnarray}
\frac{\partial}{\partial t} 
\left[\frac{\delta x(t)}{\delta \eta_m(t')} \right]
&=& [F'(x(t))
+ \sum_n G'_n(x(t)) \eta_n(t)] \frac{\delta x(t)}{\delta \eta_m(t')}.
\end{eqnarray}
The formal solution of Eq. (A10) with the initial condition:
\begin{eqnarray}
\frac{\partial}{\partial t} 
\left[\frac{\delta x(t)}{\delta \eta_m(t')} \right]&=& G_m(x(t')),
\end{eqnarray}
is given by
\begin{eqnarray}
\frac{\delta x(t)}{\delta \eta_m(t')}&=& G_m(x(t')) 
\exp\left[ \int_{t'}^{t}\:ds \: [F'(x(s)) + \sum_n G'_n(x(s)) \eta_n(x(s))]
\right].
\end{eqnarray}
The derivative of $G_m(x(t))$ with respect to $t$ becomes 
\begin{eqnarray}
\frac{d}{dt} G_m(x(t)) &=& G'_m(x(t)) \frac{d x(t)}{dt}, 
\\
&=& G'_m(x(t)) [F(x(t))+\sum_n G_n(x(t)) \eta_n(t)].
\end{eqnarray}
The integral of Eq. (A14) yields
\begin{eqnarray}
G_m(x(t')) &=& G_m(x(t)) 
\exp \left[ \int_{t}^{t'} \:ds \:\left( \frac{G'_m(x(s))}{G_m(x(s))} \right)
[F(x(s))+ \sum_n G_n(x(s)) \eta_n(x(s))] \right]. \nonumber \\ && 
\end{eqnarray}
Substituting Eqs. (A12) and (A15) to Eq. (A9), we get
\begin{eqnarray}
\frac{\delta x(t)}{\delta \eta_m(t')}
&\simeq & G_m(x(t)) \exp\left( \int_{t'}^{t} \:ds \:
\left[ F'(x(s))-\frac{G'_m(x(s))}{G_m(x(s))}F(x(s)) \right] \right),
\end{eqnarray}
where contributions from terms including $\eta_n$ in Eq.(A12) and (A15) 
are neglected.
Combining Eq. (A7) with Eq. (A16), we get
\begin{eqnarray}
&&\langle \eta_m(t) \delta(x(t)-x) \rangle
= - \frac{\partial}{\partial x} G_m(x(t)) 
\int_0^t dt' \nonumber \\
&&\times \left[c_{mm} \left< \delta(x(t)-x)
\exp\left(\int_{t'}^{t} \:ds \left[F'(x(s)) 
- \frac{G'_m(x(s))}{G_m(x(s))} F(x(s))\right] \right) \right> \right]. 
\nonumber \\ && 
\end{eqnarray}
By using the decoupling approximation given by
\begin{eqnarray}
&&\left< \delta(x(t)-x)
\exp\left(\int_{t'}^{t} \:ds \left[F'(x(s)) 
- \frac{G'_m(x(s))}{G_m(x(s))} F(x(s))\right] \right) \right> 
\\
&\simeq &
\left< \delta(  x(t)-x) \right>
\exp \left(\left<  \left[F'(x(t)) 
- \frac{G'_m(x(t))}{G_m(x(t))}F(x(t))\right] \right> (t-t') \right),
\end{eqnarray}
we get
\begin{eqnarray}
\langle \eta_m(t) \delta(x(t)-x) \rangle
&\simeq & 
\tilde{D}_m
\frac{\partial}{\partial x}G_m(x) p(x,t),
\end{eqnarray}
with
\begin{eqnarray}
\tilde{D}_m &=& \frac{D_m}{[1-\tau_m 
(\langle F' \rangle- \langle F G'_m/G_m \rangle)]},
\end{eqnarray}
where $F'=dF/dx$ and $G'_m=dG_m/dx$, and the bracket 
$\langle \cdot \rangle $ expresses the average over $p(x,t)$.
Substituting Eq. (A20) to Eq. (A4), we finally
get the effective FPE given by
\begin{eqnarray}
\frac{\partial}{\partial t}p(x,t) 
&=& - \frac{\partial }{\partial x}F(x)p(x,t)
+ \sum_{m} \tilde{D}_m \frac{\partial }{\partial x}G_m(x)
\frac{\partial }{\partial x}G_m(x) p(x,t),
\end{eqnarray}
from which we get the effective Langevin equation given by
\begin{eqnarray}
\frac{dx}{dt}\!\!&=&\!\! F(x) 
+ \sum_m \sqrt{2 \tilde{D}_m} G_m(x) \xi_m(t),
\end{eqnarray}
with
\begin{eqnarray}
\langle \xi_m(t) \rangle &=& 0, 
\\
\langle \xi_m(t) \xi_n(t') \rangle
&=& \delta_{mn} \delta(t-t').
\end{eqnarray}

\newpage

{\it Table 1} A comparison among various methods
yielding the effective equation given by 
$\dot{x}= \tilde{F}(x)+ \sqrt{2 \tilde{D}_0}\xi_0(t) 
+ \sqrt{2 \tilde{D}_1} G(x) \xi_1(t)$
for the Langevin model:
$\dot{x}= F(x)+ \eta_0(t) + G(x) \eta_1(t)$
subjected to colored noise $\eta_m$ with the relaxation time
$\tau_m$ ($m=0,1$) given by Eqs. (1) and (2);
the bracket $\langle \cdot \rangle$ denotes
the average over $p(x,t)$:
$F_s=F(x_s)$, $F'_s=F'(x_s)$ {\it etc.}
for the stationary $x_s$ (see text).
\begin{center}
\begin{tabular}{|c|c|c|c|}
\hline
$\tilde{F}$ & $\tilde{D}_0$ & $\tilde{D}_1$ & method \\ \hline \hline
$F$ & $D_0/(1-\tau_0 \langle F' \rangle)$ 
& $D_1/[1-\tau_1(\langle F'\rangle-\langle F G' / G \rangle )]$ 
& ${\rm FIM}^{a)}$ \\
$F$ & $D_0/(1-\tau_0 F'_s)$ 
& $D_1/[1-\tau_1(F'_s -F_s G'_s/G_s)]$ 
& ${\rm FIM}^{b)}$ \\
$F$ & $-$ & $D_1/[1-\tau_1(F'-FG'/G)]$ 
& ${\rm CV}^{c)} $ \\
$F/[1-\tau_1(F'-F G'/G)]$ & $-$ & $D_1/[1-\tau_1(F'-F G'/G)]^2 $
&  ${\rm UCNA} ^{d)}$ \\
$F/(1-\tau_0 F')$ & $D_0/(1-\tau_0 F')^2$ 
& $-$ & ${\rm UCNA}^{e)}$ \\
$F/(1-\tau_0 F')$ & $D_0/(1-\tau_0 F')^2$ 
& $-$ & ${\rm FIM}^{f)}$ \\
$F$ & $D_0/(1-\tau_0 \langle F' \rangle)$ 
& $-$ 
& ${\rm MM}^{g)}$ \\

\hline
\end{tabular}
\end{center}

\noindent
(a) the functional-integral method (FIM; present study).

\noindent
(b) FIM (Ref. \cite{Liang04,Wu07}).

\noindent
(c) a change of variable (CV)
for multiplicative colored noise only (after Ref. \cite{Risken96}). 

\noindent
(d) UCNA calculation for multiplicative colored noise only
(Ref. \cite{Jung87}).

\noindent
(e) UCNA calculation for additive colored noise only
(Ref. \cite{Jung87}).

\noindent
(f) FIM for additive colored noise only
(Ref. \cite{Fuentes02}).

\noindent
(g) moment method (MM) for additive colored noise only
(Ref. \cite{Hasegawa07b}).

%\end{document}

\newpage
\begin{figure}
%\begin{center}
%\includegraphics[keepaspectratio=true,width=100mm]{fig1.eps}
%\end{center}
\caption{
%Fig C.
%(Color online)
The ratio of $\tilde{D}_1/D_1$ vs. $\tau_1$ 
calculated by the FIM (solid curve), 
the approximation (APP) given by Eq. (27) 
(dashed curve)
and the difference of (FIM- APP)$ \times $ 10 (chain curve) with
$I=0.5$, $\lambda=1.0$, $D_0=0.01$, $D_1=0.2$ and $\tau_0=0.01$.
}
\label{fig1}
\end{figure}

\begin{figure}
\begin{center}
\end{center}
\caption{
%Fig C.
(Color online)
The stationary distributions $p(x)$ for
$\tau_1=0.1$ [(a), (b) ],
$\tau_1=1.0$ [(c), (d) ] and
$\tau_1=5.0$ [(e), (f)] calculated by the FIM (solid curves),
DS (dashed curves),
the approximation given by Eq. (27) (APP; chain curves) and
in the white-noise limit (WN; double chain curves)
with $I=0.5$, $\lambda=1.0$, $D_0=0.01$, $D_1=0.2$ and $\tau_0=0.01$:
(a), (c) and (e) are in normal scale,
and (b), (d) and (f) are in log scale. 
}
\label{fig2}
\end{figure}

\begin{figure}
\begin{center}
\end{center}
\caption{
%Fig C.
(Color online)
The response of $\mu(t)$ and $\gamma(t)$ to pulse input for
$\tau_1=0.1$ [(a), (b) ],
$\tau_1=1.0$ [(c), (d) ] and
$\tau_1=5.0$ [(e), (f)] calculated by the FIM (solid curves),
DS (dashed curves),
the approximation (APP) given by Eq. (27) (chain curves) and
in the white-noise limit (double chain curves)
with $\lambda=1.0$, $D_0=0.01$, $D_1=0.2$ and $\tau_0=0.01$:
$\mu(t)$ is plotted in (a), (c) and (f),
$\gamma(t)$ in (b), (d) and (f), and 
a pulse input $I(t)$ given by Eq. (35) is shown
at bottoms of (a), (c) and (e). 
The ordinate of (f) is different from those of (b) and (d):
$\gamma(t)$ in the white-noise limit (double-chain curve)
is multiplied by a factor of 1/5.
}
\label{fig3}
\end{figure}

\begin{figure}
\begin{center}
\end{center}
\caption{
%Fig C.
(Color online)
The response of $\mu(t)$ and $\gamma(t)$ to sinusoidal input for
$\tau_1=0.1$ [(a), (b) ],
$\tau_1=1.0$ [(c), (d) ] and
$\tau_1=5.0$ [(e), (f)] calculated by the FIM (solid curves),
DS (dashed curves),
the approximation given by Eq. (27) (chain curves) and
in the white-noise limit (double chain curves)
with $\lambda=1.0$, $D_0=0.01$, $D_1=0.2$ and $\tau_0=0.01$:
$\mu(t)$ is plotted in (a), (c) and (f),
$\gamma(t)$ in (b), (d) and (f), and 
a sinusoidal input $I(t)$ given by Eq. (38) is shown
at bottoms of (a), (c) and (e). 
The ordinate of (f) is different from those of (b) and (d):
$\gamma(t)$ in the white-noise limit (double-chain curve)
is multiplied by a factor of 1/5.
}
\label{fig4}
\end{figure}

\begin{figure}
\begin{center}
\end{center}
\caption{
%Fig C.
(Color online)
The ratio of $\tilde{D}_m/D_m$ vs. $\tau_1\:(=\tau_2/10)$ ($m=1,2$) 
calculated by the FIM ($m=1$, solid curve; $m=2$, dashed curve) 
and the difference of (FIM- APP)$ \times $ 50 
($m=1$, chain curve; $m=2$, double-chain curve) 
with $\lambda=1.0$, $D_0=0.01$, $D_1=0.1$, $D_2=0.2$ and $\tau_0=0.01$,
results of the APP being given by $\tilde{D}_m^{APP}$ in Eq. (59).
}
\label{fig5}
\end{figure}

\begin{figure}
\begin{center}
\end{center}
\caption{
%Fig C.
(Color online)
The stationary distributions $p(x)$ for
$\tau_1=0.05$, $\tau_2=0.5$ [(a), (b) ],
$\tau_1=0.5$, $\tau_2=5.0$ [(c), (d) ] and
$\tau_1=5.0$, $\tau_2=50.0$ [(e), (f)] 
calculated by the FIM (solid curves),
DS (dashed curves),
the approximation given by Eq. (59) (chain curves) and
in the white-noise limit (double chain curves)
with $I=0.5$, $\lambda=1.0$, $D_0=0.01$, $D_1=0.1$, 
$D_2=0.2$ and $\tau_0=0.01$:
(a), (c) and (e) are in normal scale,
and (b), (d) and (f) are in log scale. 
}
\label{fig6}
\end{figure}

\begin{figure}
\begin{center}
\end{center}
\caption{
%Fig C.
(Color online)
The response of $\mu(t)$ and $\gamma(t)$ to pulse input for
$\tau_1=0.05$, $\tau_2=0.5$ [(a), (b) ],
$\tau_1=0.1$, $\tau_2=1.0$ [(c), (d) ] and
$\tau_1=0.5$, $\tau_2=5.0$ [(e), (f)] calculated by the FIM (solid curves),
DS (dashed curves),
the approximation given by Eq. (59) (chain curves) and
in the white-noise limit (double chain curves)
with $\lambda=1.0$, $D_0=0.01$, $D_1=0.1$, $D_2=0.2$ and $\tau_0=0.01$:
$\mu(t)$ is plotted in (a), (c) and (f),
$\gamma(t)$ in (b), (d) and (f), and 
an input pulse $I(t)$ given by Eq. (35) is shown
at bottoms of (a), (c) and (e). 
The ordinate of (f) is different from those of (b and (d):
$\gamma(t)$ in the white-noise limit (double-chain curve)
is multiplied by a factor of 1/5.
}
\label{fig7}
\end{figure}

\begin{figure}
\begin{center}
\end{center}
\caption{
%Fig D.
(Color online)
Time courses of $\mu(t)$ (a) and $\gamma$ (b) for a pulse input 
and those of $\mu(t)$ (c) and $\gamma$ (d) for a sinusoidal input,
calculated by the FIM (solid curves), DS (dashed curves)
and the UCNA (chain curves) for $\lambda=1.0$, 
$D_1=0.2$ and $\tau_1=1.0$:
pulse and sinusoidal inputs $I(t)$ given by Eqs. (35) and (38),
respectively, are shown at the bottoms of (a) and (c). 
}
\label{fig8}
\end{figure}

\end{document}